\documentclass[%
 reprint,%reprint
 superscriptaddress,
 amsmath,amssymb,
 aps,
 pra,
]{revtex4-1}

\usepackage[colorlinks=true,citecolor=blue]{hyperref}
\usepackage{graphicx}% Include figure files
\usepackage{dcolumn}% Align table columns on decimal point
\usepackage{bm}% bold math
\usepackage[dvipsnames]{xcolor}
\usepackage[normalem]{ulem}
\usepackage{chngcntr}
\begin{document}

\preprint{APS/123-QED}

%---------------------------------------- Title and Authors ----------------------------------------
%===================================================================================================

\title{Giant magnetocaloric effect driven by indirect exchange in magnetic multilayers}

\author{D.~M.~Polishchuk}
\email{dpol@kth.se.}
\affiliation{Nanostructure Physics, Royal Institute of Technology, Stockholm, Sweden}%
\affiliation{Institute of Magnetism, NAS of Ukraine and MES of Ukraine, Kyiv, Ukraine}

\author{Yu.~O.~Tykhonenko-Polishchuk}
\affiliation{Nanostructure Physics, Royal Institute of Technology, Stockholm, Sweden}
\affiliation{Institute of Magnetism, NAS of Ukraine and MES of Ukraine, Kyiv, Ukraine}

\author{E.~Holmgren}
\affiliation{Nanostructure Physics, Royal Institute of Technology, Stockholm, Sweden}

\author{A.~F.~Kravets}
\affiliation{Nanostructure Physics, Royal Institute of Technology, Stockholm, Sweden}
\affiliation{Institute of Magnetism, NAS of Ukraine and MES of Ukraine, Kyiv, Ukraine}

\author{A.~I.~Tovstolytkin}
\affiliation{Institute of Magnetism, NAS of Ukraine and MES of Ukraine, Kyiv, Ukraine}
 
\author{V.~Korenivski}%
\affiliation{Nanostructure Physics, Royal Institute of Technology, Stockholm, Sweden}%

\date{\today}% It is always \today, today,
             %  but any date may be explicitly specified

\begin{abstract}
%===================================================================================================

Indirect exchange coupling in magnetic multilayers, also known as the Ruderman-Kittel-Kasuya-Yosida (RKKY) interaction, is highly effective in controlling the interlayer alignment of the magnetization. This coupling is typically fixed at the stage of the multilayer fabrication and does not allow ex-situ control needed for device applications. In addition to the orientational control, it is highly desirable to also control the magnitude of the \emph{intralayer} magnetization, ideally switch it on/off by switching the relevant RKKY coupling. Here we demonstrate a magnetic multilayer material, incorporating thermally- as well as field-controlled RKKY exchange, focused on to a dilute ferromagnetic alloy layer and driving it though it’s Curie transition. Such on/off magnetization switching of a thin ferromagnet, performed repeatably and fully reproducibly within a low-field sweep, results in a giant magnetocaloric effect, with the estimated isothermal entropy change of $\Delta S \approx -10$~mJ~cm$^{-3}$~K$^{-1}$ under an external field of $\sim$10~mT, which greatly exceeds the performance of the best rare-earth based materials used in the adiabatic-demagnetization refrigeration systems.

%\begin{description}
%\item[Usage]
%Secondary publications and information retrieval purposes.
%\item[PACS numbers]
%May be entered using the \verb+\pacs{#1}+ command.
%\item[Structure]
%You may use the \texttt{description} environment to structure your abstract;
%use the optional argument of the \verb+\item+ command to give the category of each item. 
%\end{description}
\end{abstract}

%\pacs{Valid PACS appear here}% PACS, the Physics and Astronomy
                             % Classification Scheme.
%\keywords{Suggested keywords}%Use showkeys class option if keyword
                              %display desired
\maketitle

%\tableofcontents

%\begin{figure}[b]
%\includegraphics{fig_1}% Here is how to import EPS art
%\caption{\label{fig:epsart} A figure caption. The figure captions are
%automatically numbered.}
%\end{figure}

%\cite{Brataas2002,Ando2011}
%\emph{static}
%$\mathbf{I}_\mathrm{s}^\mathrm{pump}$
%$d_\text{Cr} =$ 1.2–5~nm
%Fig.~\ref{fig_1}
%\eqref{Jrkky}

\section{Introduction}

Advances in the field of nanostructuring of magnetic materials have led to a number of important research discoveries with subsequent device demonstrations, such as the giant and tunneling magnetoresistance in thin-film multilayers (sensors in magnetic storage, memory elements, rf oscillators, etc.) \cite{Baibich1988,Binasch1989,Moodera1995}, perpendicular magnetic anisotropy in thin film particles (high-density storage elements down to $\sim$10~nm in lateral size) \cite{Carcia1985,Weller2000}, fast-moving domain walls in magnetic nanowires (domain-wall racetrack memory) \cite{Parkin2008,Parkin2015}, magnetic meta-materials (arrays of nano-objects; magnonic crystals, artificial spin ice, etc.) \cite{Kruglyak2010,Wang2006}, exotic spin-vortex states in single and stacked nanoparticles \cite{Shinjo2000,Cherepov2012}, etc. This variety of new materials and phenomena stems from reduced size effects invoked by nanostructuring the relevant surface and interface interactions, determining their magnetic properties and device functionality. The now classical example of such an interface effect is the Ruderman-Kittel-Kasuya-Yosida (RKKY) interaction in magnetic multilayers \cite{Gruenberg1986}. The RKKY interaction is an indirect exchange coupling oscillating in sign with $\sim$1~nm periodicity and rapidly decaying in magnitude as nonmagnetic spacing between two interacting magnetic interfaces is increased \cite{Parkin1990}. It has recently been shown that RKKY exchange, normally insensitive to external control, can undergo a ferromagnetic-to-antiferromagnetic transition in response to a change in temperature \cite{Polishchuk2017,Polishchuk2017prb} or applied electric field \cite{Newhouse2017}.

Here we propose that the magnetocaloric effect (MCE), defined as an isothermal entropy change or an adiabatic temperature change under an applied magnetic field, can greatly benefit from enhancing the applied field strength by the intrinsic RKKY in the nanostructure. This is in contrast to the conventional approach of using nanostructuring for tailoring the nano-material's magnetic properties, such as the direct exchange and anisotropy, aimed at enhancing the MCE in low magnetic fields, adjusting its operating temperature, or suppressing unwanted hysteresis losses \cite{Skomski2008,Skomski2010,Mukherjee2009,Michalski2012}. The conventional approach often yields only minor improvements in the magnetocaloric properties, in particular due to the relatively low energy of magnetic anisotropy compared to that of thermal fluctuations near room temperature. Here, we show that field- and temperature-control of the indirect RKKY exchange, with its action pin-pointed at specific interfaces in a multilayer, can yield greatly enhanced MCE. We demonstrate a multilayer design capable of switching between different entropy states, controlled by a directional switching of RKKY under a weak applied field. The observed RKKY-driven phase transition in a thin dilute 3d-ferromagnetic alloy layer indicates a greatly enhanced isothermal entropy change per unit field, much larger than that in the rare-earth based materials with the highest MCE response (cf. Ref.~\onlinecite{Gsch2005}).

\section{Experiment and samples details}

The multilayer samples were grown onto Ar pre-etched Si (100) substrates at room temperature using a dc magnetron sputtering system (AJA Inc.). Layers of dilute Fe$_x$Cr$_{100-x}$ binary alloys of varied composition were deposited using co-sputtering from separate Fe and Cr targets. The alloy composition was varied by setting the deposition rates of the individual Fe and Cr components based on calibrations obtained by thickness profilometry. The magnetic properties were characterized using longitudinal magneto-optical Kerr effect (MOKE) measurements in the temperature range of 77~K--460~K using a MOKE setup equipped with an optical cryostat (by Oxford Instruments). The room-temperature magnetic properties were characterized using a vibrating-sample magnetometer (VSM, by Lakeshore Cryogenics) and ferromagnetic resonance (FMR, X-band ELEXSYS E500 spectrometer by Bruker).

\section{Results}

\subsection{Isothermal entropy change enhanced by RKKY exchange}

Magnetic susceptibility of a ferromagnet steeply rises as one approaches its Curie temperature, $T_\mathrm{C}$. This leads to enhanced MCE, since an applied magnetic field effectively suppresses critical spin fluctuations associated with magnetic entropy. Such critical-point behavior can be utilized in different ways in magnetic multilayers. In particular, in a trilayer F$_1$/f/F$_2$ \cite{Andersson2010,Kravets2012}, where a low-$T_\mathrm{C}$ spacer (f) mediates exchange between two high-$T_\mathrm{C}$ ferromagnets (F$_1$ and F$_2$), the outer ferromagnets exert a strong magnetic proximity effect on f due to the direct exchange coupling at the interfaces. In the vicinity of the spacer’s $T_\mathrm{C}$, the parallel-antiparallel switching of the magnetic moments of F$_1$ and F$_2$ has recently been predicted to yield a strong entropy change in the system ($\Delta S \sim -1$~mJ~cm$^{-3}$~K$^{-1}$) \cite{Fraerman2015}. This difference in magnetic entropy between the parallel (P) and antiparallel (AP) orientations was estimated via the proximity effect on the spacer from the strongly ferromagnetic outer layers, with the contributions from the two interfaces expected to add up or cancel out in the  P or AP states of the trilayer, respectively. Our extensive studies have shown, however, that the \emph{direct exchange} across the spacer is too strong and never cancels out for realistic material parameters~\cite{Kravets2012,Kravets2015}.

The design we propose uses \emph{indirect RKKY exchange} through a thin nonmagnetic layer N incorporated at the two F/N/f interfaces, and has the following unique advantages. First, the sign and strength of the RKKY interaction are well defined by the thickness of N, even for dilute ferromagnetic alloys, due to the sharp oscillatory character of RKKY, with the period in the thickness of N of about 1~nm. This affords a great flexibility in the multilayer design in terms of matching in-phase and/or out-of-phase RKKY contributions from the opposing interfaces at a given location within the spacer. Secondly, RKKY-induced magnetic biasing can be made sufficiently strong. For example, an increase of a Ni layer’s Curie temperature by $\sim$20~K has been reported when coupled to a Co layer by RKKY \cite{Bayreuther1996,Bovensiepen1998}. Significantly larger changes in magnitude as well as sign of RKKY obtained by varying temperature were demonstrated for multilayers based on dilute ferromagnetic alloys \cite{Polishchuk2017,Polishchuk2017prb}.

%##################### Figure 1 #####################
%===================================================
\begin{figure}[t]
\includegraphics[width=8cm]{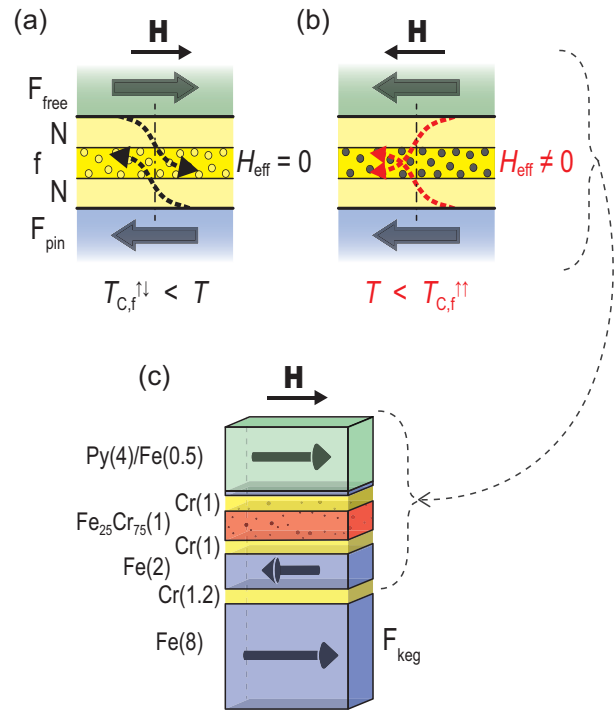}
\caption{(a), (b) Illustration of field-driven parallel-antiparallel (P-AP) magnetization switching in magnetic multilayers with gradient nonmagnetic/weakly-magnetic/ nonmagnetic (N/f/N) spacer and strong indirect exchange bias. Gradient spacer, N/f/N, is designed such that thin inner layer f is affected by two RKKY contributions, either adding or subtracting depending on mutual orientation of outer F$_\mathrm{free}$ and F$_\mathrm{pin}$ ferromagnets. (c) Fabricated multilayer layout, F$_\mathrm{free}$/N/f/N/F$_\mathrm{pin}$/N/F$_\mathrm{keg}$. Arrows indicate magnetization of ferromagnetic layers F$_\mathrm{free}$, F$_\mathrm{pin}$ and F$_\mathrm{keg}$ in weak magnetic field $\mathbf{H}$.}
\label{fig_1}
\end{figure}
%===================================================

Strong RKKY biasing by the outer ferromagnets, focused at a sufficiently thin weakly ferromagnetic inner spacer layer, f, is used in our design to substantially alter the magnetic order in f in the vicinity of its Curie temperature, $T_\mathrm{C,f}$ [Fig.~\ref{fig_1}(a,b)]. The thickness of f should be small enough not to significantly exceed the RKKY-exchange penetration depth (typically a few monolayers). The parallel alignment of F$_1$ and F$_2$, with additive RKKY contributions at the inner-spacer, results in strong RKKY-exchange biasing of f, ordering it magnetically, significantly increasing its Curie temperature, $T_\mathrm{C,f}^{\uparrow \uparrow}$, and quenching its magnetic entropy [Fig.~\ref{fig_1}(b)]. The antiparallel alignment of F$_1$ and F$_2$, on the other hand, has the two RKKY contributions at f directed in opposition, canceling out the total RKKY exchange, which manifests as a much lower effective Curie temperature of f,  $T_\mathrm{C,f}^{\uparrow \downarrow} < T_\mathrm{C,f}^{\uparrow \uparrow}$ [Fig.~\ref{fig_1}(a)]. This interplay is a function of the biasing strength and, for the temperature range $T_\mathrm{C,f}^{\uparrow \downarrow} < T < T_\mathrm{C,f}^{\uparrow \uparrow}$, switching of one of the outer layers (F$_1$ or F$_2$) switches the RKKY and effectively drives the magnetic phase transition in f, as illustrated in Fig.~\ref{fig_1}(a,b). Such a strong magnetic phase transition, with the RKKY exchange rather than only the applied field switched on/off, takes place at a constant temperature, yielding a greatly-enhanced isothermal entropy change. 

%##################### Figure 2 #####################
%===================================================
\begin{figure*}%[b]
\includegraphics[width=14cm]{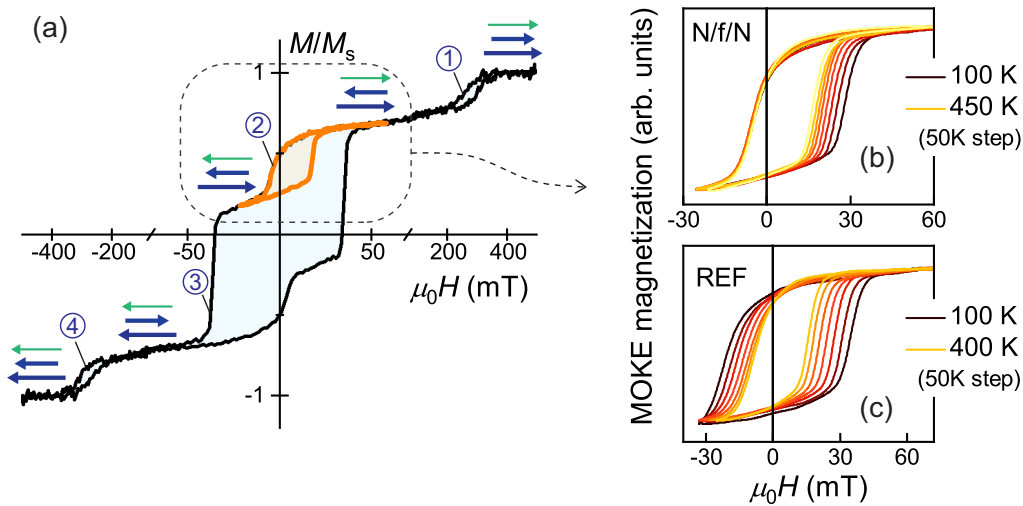}
\caption{(a) Full $M$-$H$ loop for multilayer with gradient spacer, N/f/N = Cr(1)/Fe$_{25}$Cr$_{75}$(1)/Cr(1), measured using VSM at room temperature. Arrows indicate magnetization of F$_\mathrm{free}$ (green), F$_\mathrm{pin}$ (short blue) and F$_\mathrm{keg}$ (long blue) during positive-to-negative field sweep. Shown in orange is the measured minor loop of F$_\mathrm{free}$. (b), (c) Temperature evolution of minor magnetization loop of F$_\mathrm{free}$ measured using MOKE for structures with (b) gradient N/f/N spacer ($t_\mathrm{f} =$ 1~nm) and (c) uniform spacer, REF = Cr(6).}
\label{fig_2}
\end{figure*}
%===================================================

\subsection{Thermomagnetic RKKY-vs-intrinsic exchange tuning: multilayer design}

The system with perhaps the strongest RKKY is the now classical Fe/Cr multilayers \cite{Baibich1988,Binasch1989}, where the strength and sign of the interlayer exchange is determined by the thickness of the Cr layer, chosen during fabrication \cite{Parkin1990}. Incorporation of thin, low-$T_\mathrm{C}$ Fe$_x$Cr$_{100-x}$ dilute-ferromagnetic alloy spacers within the Cr layers can make the RKKY in the system strongly temperature dependent in the desired operating range \cite{Polishchuk2017,Polishchuk2017prb}. Here we use this control mechanism and show how the RKKY within the spacer can be doubled or cancelled out essentially completely by a sweep of a 10~mT range external magnetic field. This results in a ferro-to-paramagnetic phase transition in the dilute ferromagnetic alloy, its full demagnetization, and consequently a giant magnetocaloric effect.

The material design principle is illustrated in Fig.~\ref{fig_1}(c). The free layer is essentially a thin layer of Permalloy, covered at the inner interface with atomically thin Fe: F$_\mathrm{free}$~= Py(4)/Fe(0.5), with the thickness in nanometers given in the parentheses. The soft Permalloy layer (Py = Ni$_{80}$Fe$_{20}$) provides low coercivity of F$_\mathrm{free}$, whereas the ultra-thin Fe(0.5) is used to increase the electron spin-polarization at the F$_\mathrm{free}$/N interface, which greatly enhances the strength of the RKKY interaction \cite{Parkin1992}. The gradient spacer N/f/N, with N = Cr(1) and f~= Fe$_{25}$Cr$_{75}$, has the thickness of the pure Cr layers (1~nm) chosen to correspond to strong antiferromagnetic RKKY (first antiferromagnetic RKKY peak). The nominally weakly ferromagnetic Fe$_{25}$Cr$_{75}$ inner spacer is well lattice matched within the Cr/Cr-Fe/Cr spacer, has near perfect miscibility of Fe in Cr, and has a suitably low bulk Curie temperature of $T_\mathrm{C} \approx$ 150~K \cite{Burke1978,Babic1980}. The intriguing thickness range for the inner spacer f is a few monolayers ($t_\mathrm{f} \sim$ 1~nm), where, as detailed in the sections that follow, the combined interfacial RKKY exchange from the two outer Fe electrodes becomes comparable in magnitude to the intrinsic exchange within the Cr-Fe alloy and can add to it (parallel F$_\mathrm{free}$ and F$_\mathrm{pin}$) or mutually subtract (antiparallel F$_\mathrm{free}$ and F$_\mathrm{pin}$), thereby driving a magnetic phase transition in the structure, with a tunable operating (Curie) point.

For clearly monitoring this thermomagnetic transition, one of the outer layers (F$_\mathrm{pin}$) ideally is fixed in its magnetization direction, which is usually done by exchange-pinning it to an antiferromagnet~\cite{Nogues1999,Nogues2005}. Such pinning, however, is known to have a strong temperature dependence for widely used metallic antiferromagnets, such as Ir$_{80}$Mn$_{20}$ or Fe$_{50}$Mn$_{50}$, and the typical pinning fields at around room-temperature are rarely above 50~mT, even for nanometer thin Fe films \cite{Nogues1999}. Our pinned layer, F$_\mathrm{pin}$, was specially designed to provide efficient, essentially temperature-independent pinning up to at least 200~mT, needed for a reliable control of the RKKY in the structure. It uses the strong antiferromagnetic-RKKY peak \cite{Parkin1991} between F$_\mathrm{pin} =$ Fe(2) and a much thicker F$_\mathrm{keg} =$ Fe(8) in a Fe(2)/Cr(1.2)/Fe(8) trilayer, hereafter refered to as the synthetic ferrimagnet (SFM).  Due to the four-fold difference in thickness, the ground state of the SFM always has the magnetization of F$_\mathrm{keg}$ aligned with the external field, with the F$_\mathrm{pin}$ moment strictly antiparallel to F$_\mathrm{keg}$ (and field), up to about 200~mT. The above specially designed RKKY- and thermally-tunable spacer, in combination with the tailored to it magnetically stiff and temperature insensitive pinned layer, make possible the exchange-enhanced magnetocaloric effect demonstrated below.

\subsection{Free layers' coercivity as probe of spacer's thermo-magnetism}

The major magnetization loop of a multilayered structure described above, shown in Fig.~\ref{fig_2}(a), exhibits clear switching for each of the strongly ferromagnetic layers during the field sweep. At fields in access of about 300~mT all layers are aligned in parallel along the field by the strong Zeeman interaction. The high-field step (1) between 300 and 200~mT corresponds to switching of the pinned layer, F$_\mathrm{pin} =$ Fe(2), where the strong antiferromagnetic RKKY in the reference SFM overcomes the Zeeman energy of F$_\mathrm{pin}$. Step (2) near zero field corresponds to switching of the free layer, returning F$_\mathrm{free}$ and F$_\mathrm{pin}$ to a mutually parallel orientation. Step (3) at an intermediate negative field corresponds to switching of the thick layer within the SFM, F$_\mathrm{keg} =$ Fe(8), which, due to the strong interlayer coupling in SFM, also switches F$_\mathrm{pin}$ -- the synthetic ferrimagnet reverses by in-phase rotation, preserving its intrinsic AP state. Step (4) at a large negative field corresponds to F$_\mathrm{pin}$ aligning with the field and the entire structure saturating opposite to the original direction. Since the coercive fields of all the ferromagnetic layers are well separated, with the pinned layer highly stable and temperature insensitive at low fields, this multilayer design enables probing the RKKY-induced thermo-magnetic transition within the specially designed gradient Cr/Cr-Fe/Cr spacer via the switching of F$_\mathrm{free}$ within the minor loop, shown in orange in Fig.~\ref{fig_2}(a).

The temperature evolution of the minor loop of F$_\mathrm{free}$ for $t_\mathrm{f} =$ 1~nm, shown in Fig.~\ref{fig_2}(b), reveals the key difference between the gradient spacer, incorporating a dilute ferromagnetic alloy layer, and a reference uniform pure-Cr spacer, sp~= Cr(6), Fig.~\ref{fig_2}(c). With decreasing temperature, the minor loop for the gradient-spacer structure displays a significant increase in the positive (right) coercivity field, corresponding to the P to AP switching, $H_\mathrm{c}^{\uparrow\uparrow}$, while the negative (left) coercive field, corresponding to the AP to P switching, $H_\mathrm{c}^{\uparrow\downarrow}$, remains unchanged. In contrast, the positive and negative $H_\mathrm{c}$ for the reference multilayer with a pure-Cr spacer have qualitatively the same variation with temperature, with both $H_\mathrm{c}$ fields increasing at low temperatures, -- the behavior expected for a free single magnetic film. We note this principal change of behavior of the coercivity for one specific transition -- AP to P switching between the free and pinned Fe layers enclosing the spacer. We note further that, generally, coercivity scales inversely with the number of thermal magnons in the system, so the absence of temperature dependence in $H_\mathrm{c}^{\uparrow\downarrow}$ but not in $H_\mathrm{c}^{\uparrow\uparrow}$ points to a magnon-saturated state (AP), as discussed in detail below.

%##################### Figure 3 #####################
%===================================================
\begin{figure}%[b]
\includegraphics[width=8.5cm]{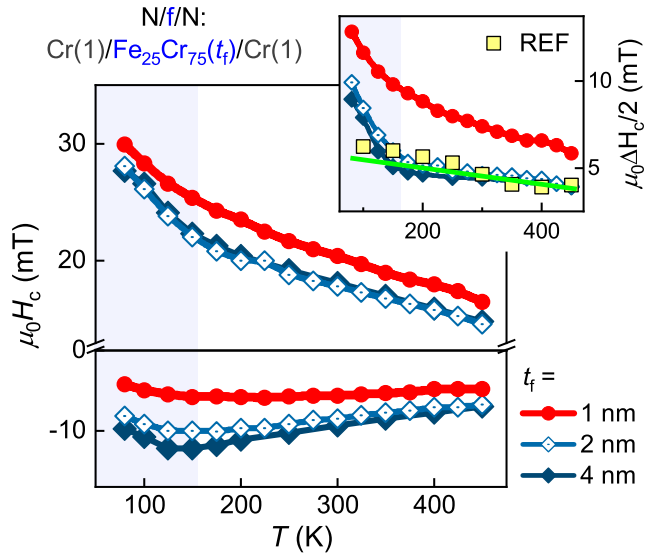}
\caption{Temperature dependence of negative and positive coercive fields $H_\mathrm{c}$ of F$_\mathrm{free}$ for thickness $t_\mathrm{f} =$ 1, 2 and 4~nm for multilayers with gradient spacer Cr(1)/Fe$_{25}$Cr$_{75}(t_\mathrm{f})$/Cr(1). Inset shows half-difference between right and left $H_\mathrm{c}$, $\Delta H_\mathrm{c}/2 = (H_\mathrm{c}^{\uparrow\uparrow} - H_\mathrm{c}^{\uparrow\downarrow})/2$, for above three thicknesses as well as for reference sample with uniform spacer Cr(6) (REF). Green linear extrapolation is guide to eye. Shaded area below 160~K indicates for magnetic-spacer f region below its nominal (bulk) Curie temperature.}
\label{fig_3}
\end{figure}
%===================================================

Figure~\ref{fig_3} shows the temperature dependence of the left and right coercive fields of F$_\mathrm{free}$ for samples with varying thickness of the weakly ferromagnetic layer, f, in the gradient spacer: $t_\mathrm{f} =$ 1, 2 and 4~nm. The thicker layers, $t_\mathrm{f} =$ 2 and 4~nm, exhibit a temperature dependence similar to that of the reference spacer, showing that for thicker f switching of the outer layers has a negligible effect on the spacer. In contrast, the sample with thin f, $t_\mathrm{f} =$ 1~nm, demonstrates a significant difference between positive and negative coercivities versus temperature, indicating a qualitative change in the thermo-magnetism of the spacer on P-AP switching in the multilayer material (red vs green and yellow in the inset to Fig.~\ref{fig_3}). 

Temperature dependence of left coercivity for $t_\mathrm{f} =$ 2 and 4~nm has a characteristic upturn near 160~K. This change is more pronounced in the half-difference between the left and right coercivities, $\Delta H_\mathrm{c}/2$, shown in the inset to Fig.~\ref{fig_3}. We ascribe this variation at lower temperatures to the conventional ferromagnetic phase transition in f into its ordered state at $T < T_\mathrm{C}^\mathrm{f} \approx $ 160~K. This transition temperature, $T_\mathrm{C}^\mathrm{f}$, is close to the known bulk Curie point of the Fe$_{25}$Cr$_{75}$ alloy, $T_\mathrm{C}^\mathrm{bulk} \approx$ 150~K. With temperature decreasing below $T_\mathrm{C}^\mathrm{f}$, magnetic ordering in f is enhanced and the interlayer coupling between the outer F$_\mathrm{free}$ and F$_\mathrm{pin}$ layers is transferred more efficiently, which shifts the $M$-$H$ loop of F$_\mathrm{free}$ toward positive fields. This interpretation is supported by additional experiments, comparing the gradient and uniform spacers (found in Supplemental Material \cite{SupplMater}).

For all samples, a slight field offset is present in the minor loop of F$_\mathrm{free}$, even when no interlayer coupling is present, such as the offset in the reference sample with uniform spacer Cr(6), Fig.~\ref{fig_2}(c). This offset is attributed to the magnetostatic N\'eel coupling between the SFM and the free layer F$_\mathrm{free}$ caused by the finite roughness of the film surface \cite{Kools1999} (see Supplemental Material \cite{SupplMater}). This offset is same for all structures studied, indicated by the green interpolation line in the inset to Fig.~\ref{fig_3}, which shows that its origin does not depend on the specifics of the spacer in the considered thickness range. In the following analysis, we subtract this spacer-independent offset, $\Delta H_\mathrm{c}^\mathrm{ms}/2$, which yields the effective coercive fields, $H_{\mathrm{c}\ast} = H_\mathrm{c} - \Delta H_\mathrm{c}^\mathrm{ms}/2$.

\subsection{Experimental evidence for RKKY-induced magnetic phase transition} 

The temperature dependence of $H_{\mathrm{c}\ast}$ for different thicknesses of the weakly magnetic layer are shown in Fig.~\ref{fig_4}. We focus on the spacer properties above its bulk Curie temperature, $T_\mathrm{C}^\mathrm{f} =$ 160~K, excluding the more complex behavior at lower temperatures irrelevant for the presented exchange-enhanced MCE effect. The left and right coercive fields ($H_\mathrm{c\ast}^{\uparrow\downarrow}$ and $H_\mathrm{c\ast}^{\uparrow\uparrow}$, defined in the inset to Fig.~\ref{fig_4}) coincide for the thicker spacers ($t_\mathrm{f} =$ 2, 4~nm) in the entire experimental temperature range (gray data points in Fig.~\ref{fig_4}). $H_\mathrm{c\ast}^{\uparrow\downarrow}$ and $H_\mathrm{c\ast}^{\uparrow\uparrow}$ differ significantly for the samples with the thinnest spacer ($t_\mathrm{f} =$ 1~nm), for which the magnitude of $H_\mathrm{c\ast}^{\uparrow\uparrow}$ is much larger and demonstrates a much stronger temperature dependence compared with that of $H_\mathrm{c\ast}^{\uparrow\downarrow}$, which essentially is temperature independent.

The pronounced asymmetry in the measured coercivity of the free layer for $t_\mathrm{f} =$ 1~nm is clearly due to the P-AP switching, designed to sum or subtract the RKKY exchange at the inner-spacer. The data suggests that with the spacer only a few atomic layers thick, the two interfacial indirect exchange profiles overlap within the spacer, ordering it ferromagnetically in the P state and fully disordering in the AP state. The temperature dependence is also quite different -- the rather strong decay of $H_\mathrm{c\ast}^{\uparrow\uparrow}(T)$ vs $T$ is the expected behavior for a conventional magnetic film, whereas the nearly constant $H_\mathrm{c\ast}^{\uparrow\downarrow}(T)$ reflects a saturation in terms of thermal magnons, which would indeed be the case for a proximal disordered spacer in the AP-state of the multilayer at all $T > T_\mathrm{C}^\mathrm{f} =$ 160~K.

The experimental data fully confirm the behavior expected from the multilayer design for a ferro-to-paramagnetic transition in the gradient spacer driven by the RKKY switching in the structure. This transition is associated with the exchange-enhanced magneto-caloric effect -- the mechanism discussed in detail in the next section.

\section{Discussion}

\subsection{Coercivity asymmetry as due to thermal activation}

Coercivity in polycrystalline films is known to be a thermally-activated process, where reverse-domain nucleation and propagation during the magnetization switching in the film proceeds via thermal activation out of local potential minima, with a morphologically determined characteristic activation energy \cite{Chikazumi2005}. Phenomenologically, the temperature dependence of the coercive field can be represented as the intrinsic coercive field reduced by thermal agitation,

%##################### Equation 1 #####################
%===================================================
\begin{equation}
H_\mathrm{c} = H_\mathrm{c0} - \Delta_{H\mathrm{c}} \exp\left(-\frac{Q}{\varepsilon k_\mathrm{B} T}\right),
\label{Hc}
\end{equation}
%===================================================

\noindent where $Q$ is the activation energy, with the associated activation temperature $T_\mathrm{a} = Q/\varepsilon k_\mathrm{B}$, $k_\mathrm{B}$ -- the Boltzmann constant, $H_\mathrm{c0}$ -- zero-temperature coercive field, $\Delta_{H\mathrm{c}}$ -- reduction in $H_\mathrm{c}$ at high temperatures ($T \gg T_\mathrm{a}$). The dimensionless parameter $\varepsilon$ is a scaling factor for the effective temperature of the free layer, which is proportional to the amount of thermal magnons external to the free layer, incoming as a magnon flux – the magnon factor.

%##################### Figure 4 #####################
%===================================================
\begin{figure}%[b]
\includegraphics[width=8.5cm]{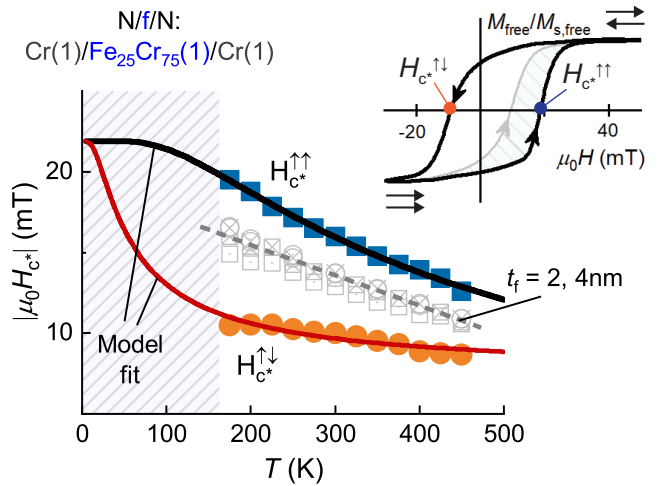}
\caption{Model analysis of temperature dependences of coercivity. Blue and orange data sets correspond, respectively, to right and left coercive fields $H_{\mathrm{c}\ast}$ of F$_\mathrm{free}$ for sample with $t_\mathrm{f} =$ 1~nm (magnetostatic offset subtracted). Gray data points correspond to samples with $t_\mathrm{f} =$ 2 and 4~nm (same within measurement uncertainty; dashed gray line is guide to eye). Solid black and orange lines are fits to experimental data using equation~\eqref{Hc}. Inset defines left ($H_{\mathrm{c}\ast}^{\uparrow\downarrow}$) and right ($H_{\mathrm{c}\ast}^{\uparrow\uparrow}$) coercive fields in hysteresis loop of F$_\mathrm{free}$.}
\label{fig_4}
\end{figure}
%===================================================

Scaling factor $\varepsilon$ in our case must depend on the mutual orientation of the outer Fe layers enclosing the gradient spacer, $\varepsilon^{\uparrow\downarrow}$ and $\varepsilon^{\uparrow\uparrow}$ for AP and P states, as that clearly results in qualitatively different $H_\mathrm{c}$ vs $T$ (different thermally-activated regimes). The transition into the regime dominated by thermal agitation takes place at $\varepsilon k_\mathrm{B} T > Q$ (or $T > T_\mathrm{a}$), where, for a fixed $Q$, an increase in temperature, ideally in the vicinity of the critical point of the spacer, greatly increases the number of thermal magnons available for agitation for the magnetization switching in the free layer. As a result, at $T \leq T_\mathrm{a}$, the free layer coercivity is expected to show a strong temperature dependence, while at $T > T_\mathrm{a}$, thermal activation due to the `external' magnon flux from the spacer is maximized and $H_\mathrm{c}$ levels off. Equation~\eqref{Hc} is a good approximation for temperatures sufficiently far from the Curie temperature of the strong ferromagnets, which is the case for our multilayers. For F$_\mathrm{free} =$ NiFe, $T_\mathrm{C}^\mathrm{free} \approx$ 800~K, so the change in the `intrinsic' magnon number in the free layer can be neglected at the operating temperature range near or below room temperature.

The temperature dependence of $H_\mathrm{c\ast}^{\uparrow\downarrow}$ and $H_\mathrm{c\ast}^{\uparrow\uparrow}$ are well fitted by equation~\eqref{Hc} in the entire measurement range, as shown by the solid lines in Fig.~\ref{fig_4}. The activation energy, $Q$, generally is a function of the local morphology and magnetic anisotropy profiles in the material and, to first order, should be temperature independent. Taking $Q$ as constant, the ratio of the effective temperatures can be extracted from the two fits through the magnon factor, $\varepsilon^{\uparrow\downarrow}/\varepsilon^{\uparrow\uparrow} \approx 7$, indicating a 7-fold enhancement in the effective temperature of the magnon bath in our system on switching from the P to AP state at a constant temperature in the operating range, which includes room temperature. The corresponding change in the number of magnons, $N_\mathrm{m}^\mathrm{inh}$, taken to scale in accordance with Bloch's law \cite{Kittel2004}, is $7^{3/2}$ or 20-fold, which is a giant change indeed for relatively low-field P-AP switching in our system. The source of this 20-fold increase in the magnon flux is the gradient spacer driven from its magnetically ordered (relatively few magnons) to fully disordered (maximum number of magnons, all Fe spins thermally agitated) by the carefully designed RKKY switching -- from \emph{constructive} to \emph{destructive} interference at the spacer's weakly magnetic core.

A quick estimate of the actually available magnons for the above thermo-magnetic effect done by counting the atomic spins in the spacer is quite informative. The model fitting in Fig.~\ref{fig_4} for the parallel state (negligible magnon flux from the spacer) yields the activation temperature of $T_\mathrm{a}^{\uparrow\uparrow} \approx$ 375~K, for which the corresponding magnon number can be estimated using Bloch's law as $N_\mathrm{m}^\mathrm{inh} \approx 2\times 10^{15}$~cm$^{-2}$. Switching into the antiparallel state (maximum magnon flux from the spacer) effectively yields, as shown above, a 7-fold reduction in the activation temperature, $T_\mathrm{a}^{\uparrow\downarrow} \approx$ 50~K, implying that the layer already has $\sim 2\times 10^{15}$~cm$^{-2}$ magnons at $T \gtrsim T_\mathrm{a}^{\uparrow\downarrow}$, whereas the Bloch's law gives a 20~times smaller magnon number for this temperature. Thus, the change in the thermal magnon population on P-AP switching in the structure, needed to explain the observed behavior, is approximately $\Delta N_\mathrm{m}^\mathrm{inh,f} \approx 2\times 10^{15}$~cm$^{-2}$. Now, taking the inner spacer as fully paramagnetic, the number of magnons it would emit is 2 per iron atom (bcc Fe has $\sim 2\mu_\mathrm{B}$~per atom). Using the relevant volume of the dilute Fe-Cr spacer, the maximum number of magnons in f (25~\%~Fe) is $N_\mathrm{m}^\mathrm{inh,f} \approx 4\times 10^{15}$~cm$^{-2}$, half of which should flow toward the free layer, providing thermo-magnetic agitation and thereby reducing its coercivity. The agreement of the estimates in this simple yet most direct comparison is excellent and supports our conclusion that the observed effect is due to the RKKY-driven ferro-to-paramagnetic transition in the gradient spacer acting as a controlled bath of thermal magnons. The effect is achieved using relatively a very low external field, of the order of 10~mT instead of the would-be-required 20~T (estimated RKKY exchange field per interface; see next section).

\subsection{Isothermal entropy change} 

The isothermal entropy change is one of the two quantitative characteristics of the MCE effect used, the other being the adiabatic temperature change. The former is often obtained indirectly via the temperature dependence of the magnetization measured in different magnetic fields and converted using the Maxwell relation. This approach for our case, however, is faced with the difficulty of separating, using conventional magnetometry, the changes in the weak magnetization of the thin MCE-active region (Fe$_{25}$Cr$_{75}$ layer) from those in the sticker, magnetically-stronger outer layers. Specifically, the changes in the magnetization of f on field sweep occur simultaneously with the magnetization switching of the much stronger free layer, which in fact drives the MCE transition in the central spacer layer via the RKKY exchange. On the other hand, we find that the well-established mean-field approach (Refs.~\onlinecite{Skomski2008} and \onlinecite{Skomski2010}) is highly suited for obtaining the isothermal entropy change for our system, once the constant of the RKKY exchange is established from the standard magnetometry analysis.

The interlayer coupling in a F$_1$/N/F$_2$ trilayer contributes to the Gibbs free energy per unit area as $-J\left(\mathbf{m}_1 \cdot \mathbf{m}_2\right)$, where $\mathbf{m}_i = \mathbf{M}_i/M_i$ is the normalized magnetic moment and $J$ is the interlayer exchange constant. With no in-plane magnetic anisotropy and equal saturation magnetizations ($M_1 = M_2 = M$) for two interacting ferromagnetic layers F$_1$ and F$_2$ of different thicknesses, $t_1 \neq t_2$, one can express $J$ through the saturation field $H_\mathbf{s}$ as follows \cite{Zhang1994}:

%##################### Equation 2 #####################
%===================================================
\begin{equation}
J = M H_\mathrm{s} \frac{t_1 t_2}{t_1 + t_2}.
\label{Jrkky}
\end{equation}
%===================================================

\noindent Here $H_\mathrm{s}$ corresponds to the effective exchange field, $H_\mathrm{ex} = H_\mathrm{s}$, acting on each layer. If one of the layers is considerably thicker (e.g., $t_1 \ll t_2$), or when it is strongly exchange biased, equation~\eqref{Jrkky} can be rewritten as $J = M H_\mathrm{s} \cdot t_1$.

Using equation~\eqref{Jrkky} we can estimate the interfacial RKKY exchange constants for our inner Fe-Cr spacer for the most interesting case of the pure-Cr layer thickness of about 1~nm (for more data and analysis see Supplemental Material \cite{SupplMater}). This straightforwardly yields $J \approx -1.0$~mJ/m$^2$, using which and taking the magnetization of the Fe$_{25}$Cr$_{75}$ dilute ferromagnetic alloy to be $M_\mathrm{f} \approx 0.25M_\mathrm{Fe}$, one obtains the effective indirect-exchange field $H_\mathrm{ex} \approx 10$~T per inner spacer's interface.

When the outer ferromagnets switch from antiparallel to parallel orientation, the total RKKY-exchange biasing field of a sufficiently thin inner spacer layer f ($\sim 1$~nm, such that the interfacial RKKY penetrates throughout the layer) changes from near perfect cancellation to $H_\mathrm{ex} \times 2 \approx 20$~T, strong enough to produce a phase transition in the highly susceptible paramagnetic layer into its magnetically ordered phase. This exchange-induced process is accompanied by a large isothermal entropy change, which can be estimated using the mean-field approximation \cite{Skomski2008}:

%##################### Equation 3 #####################
%===================================================
\begin{equation}
\Delta s = \frac{1}{2} \left(-\frac{m_\mathrm{s}\mu_\mathrm{B}\mu_0 H_\mathrm{eff}}{k_\mathrm{B} T}\right)^2 \left[\mathrm{in~units~of~} k_\mathrm{B}/\mathrm{atom}\right],
\label{s-change}
\end{equation}
%===================================================

\noindent Here $\mu_\mathrm{B}$ is the Bohr magneton, $m_\mathrm{s} \approx 2$ -- atomic spin number of Fe, $H_\mathrm{eff}$ -- effective field acting on the atomic spins. Taking $H_\mathrm{eff} = 20$~T, the entropy change $\Delta s$ in the thin Fe$_{25}$Cr$_{75}$ layer is about 0.02~$k_\mathrm{B}$/Fe. This value is consistent with those estimated for similar structures in the theoretical studies \cite{Skomski2008,Skomski2010}. Using the concentration of Fe atoms in the alloy of $n \approx 2 \cdot 10^{22}$~Fe/cm$^3$, the total entropy change in f becomes $\Delta S = n k_\mathrm{B}\Delta s \approx -10$~mJ~cm$^{-3}$~K$^{-1}$ (or $-1.4$~mJ~g$^{-1}$~K$^{-1}$, taking 7.33~g~cm$^{-3}$ for the known density of the Fe$_{25}$Cr$_{75}$ alloy). This value is about the record MCE magnitude level for the most advanced bulk magnetocaloric materials \cite{Gsch2005}, conventionally obtained by applying an external field of several tesla. In our case of a specially nanostructured material, however, a field of only a few tens of mT is needed to drive the comparable magnetic entropy change. For example, the values of the isothermal entropy change in the field range 0--10~mT for room-temperature-MCE in Gd \cite{Dankov1998} and the record-braking MCE in MnAs \cite{Wada2001} are 0.022~mJ~cm$^{-3}$~K$^{-1}$ and 0.17~mJ~cm$^{-3}$~K$^{-1}$, respectively. The observed RKKY-enhanced low-field MCE effect is two to three orders of magnitude larger than that reported for the best MCE materials. The active region in our case is about 10~\% of the full multilayer volume so a practical estimate should scale down the effect by an order of magnitude to account for the heat capacity of the `passive' layers in the stack. For example, the adiabatic temperature change $\Delta T_\mathrm{ad}$, which can be calculated as $\Delta T_\mathrm{ad} = T\Delta S/C_\mathrm{p}$, would be about 1~K for only the active layer and about 0.1~K for the full stack ($C_\mathrm{p} =$ 450~mJ~g$^{-1}$~K$^{-1}$ for both Fe and Cr). Optimizing the structure by scaling down the thick biasing layers and layering the stack should bring these two limits closer together as well as provide larger RKKY-MCE in absolute terms, potentially making it attractive for micro-coolers and heat-exchangers.

\section{Conclusions}

We demonstrate a magnetic multilayer material where indirect interlayer exchange can be switched on or off and is used to drive a specially designed dilute ferromagnetic spacer between its magnetically ordered and disordered states, which results in a giant isothermal entropy change. In its disordered state, the spacer acts as an excess magnon-bath to the free layer, greatly increasing the effective magnon temperature, reducing the free layer's coercive field used as a local probe of the thermo-magnetic transition. The indirect exchange can be toggled by external fields as small as a few milli-Tesla, determined entirely by the coercive properties of the free layer. The amplification of the effective field acting on the spacer (Zeeman into exchange) is one thousand fold, $\sim$10~mT to the estimated $\sim$20~T, which result in a commensurate entropy-per-field enhancement. 

We calculate the associated isothermal entropy change of the spacer to be $\Delta S \approx -10$~mJ~cm$^{-3}$~K$^{-1}$, a value comparable to bulk magnetocaloric materials based on rare earth materials. These results showcase how indirect exchange bias can be used to achieve large magnetocaloric effects in systems of solely transition metals, without expensive and environmentally unfriendly rare earths. We believe such systems have a great potential for small to micro-sized refrigerators, heat-exchangers, cooled micro- and nano-sensors as well as low-field tunable spin-wave emitters and other novel applications, especially those dependent on drastic miniaturization. 

%------------------------------ Acknowledgements -----------------------------
%===================================================================================================

\begin{acknowledgments}
Support from the Swedish Stiftelse Olle Engkvist Byggm\"astare, the Swedish Research Council (VR Grant No.~2014-4548), and the Swedish Institute Visby Programme are gratefully acknowledged. The work is partially supported by the National Academy of Sciences of Ukraine, proj. No. 0114U000092 and 0118U003265.
\end{acknowledgments}

\bibliography{Refs}% Produces the bibliography via BibTeX.

\section*{Supplemental material}

\subsection*{Note 1. Exchange bias due to interlayer coupling: uniform spacer case}

In a multilayer system F$_\mathrm{free}$/f/F$_\mathrm{pin}$ with uniform low-$T_\mathrm{C}$ spacer f, the position in field of the minor magnetization loop of F$_\mathrm{free}$ depends on whether or not f transmits exchange between F$_\mathrm{free}$ and F$_\mathrm{pin}$. When f is paramagnetic, there is no interlayer coupling and the minor loop of F$_\mathrm{free}$ is centered about zero field. When f is ferromagnetic, however, the minor loop is offset in field by the direct exchange through f [Kravets et al. \textit{Phys. Rev. B} \textbf{86}, 214413 (2012)].

%##################### Figure A #####################
%===================================================
\begin{figure}%[!b]
\includegraphics[width=8.5cm]{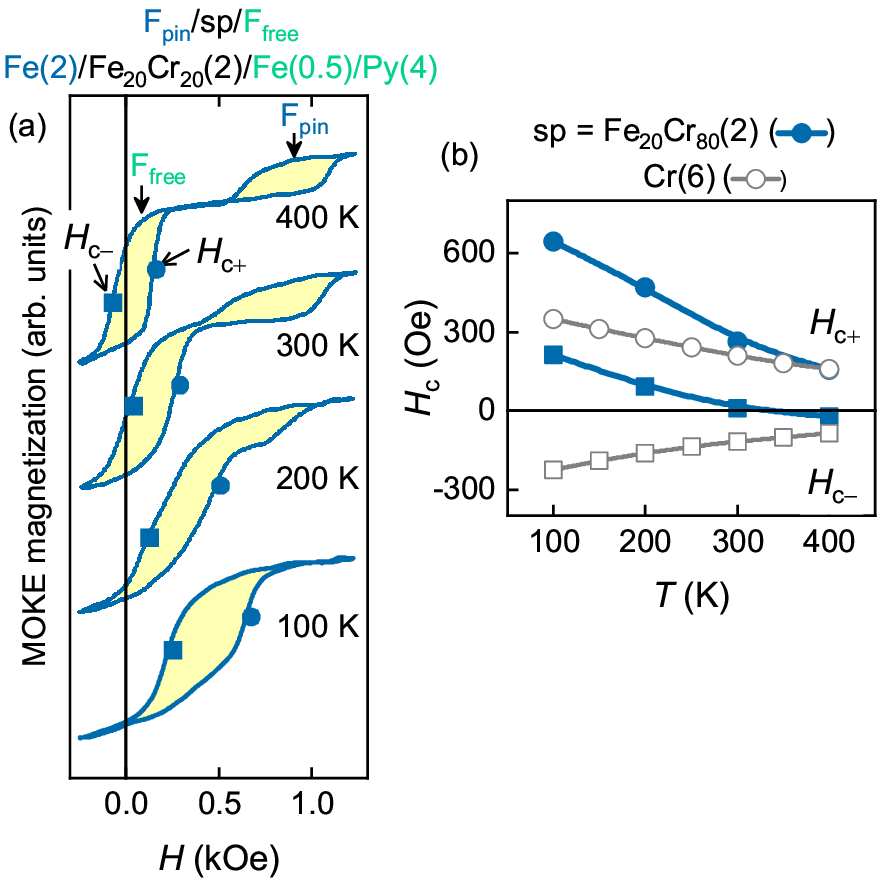}\\
\textbf{Figure A.} (a) MOKE magnetization loops for structure with continuous spacer Fe$_{20}$Cr$_{80}$(2) for various temperatures. (b) Temperature dependence of left ($H_\mathrm{c-}$) and right ($H_\mathrm{c+}$) coercive fields, as defined in panel (a), for structures with uniform spacer Fe$_{20}$Cr$_{80}$(2) and nonmagnetic spacer Cr(6).
\label{fig_A}
\end{figure}
%===================================================

We use a series of reference samples with uniformly alloyed spacers, f = Fe$_{20}$Cr$_{80}$(2~nm), in order to verify and quantify the direct interlayer exchange, as opposed to the RKKY exchange for the main gradient-spacer sample series, between F$_\mathrm{pin}$ and F$_\mathrm{free}$. MOKE magnetization loops shown in Fig.~A(a) include the minor loops of F$_\mathrm{free}$~= Fe(0.5)/Py(4) and F$_\mathrm{pin}$~= Fe(2), which are well separated at 400~K, indicating a fully decoupled state of the F$_\mathrm{free}$/f/F$_\mathrm{pin}$ trilayer. With decreasing temperature, these minor loops gradually merge, manifesting as a single exchange-offset loop at 100~K. The temperature dependence of the left ($H_\mathrm{c-}$) and right ($H_\mathrm{c+}$) coercive fields shown in (b), defined as the fields of the steepest $M$-$H$ slope [squares and circles in (a), respectively], reflect the above discussed configuration of direct-exchange pinning of F$_\mathrm{free}$. At high temperature, the $H_\mathrm{c}(T)$ data for the uniform-spacer and the pure-Cr-spacer samples [at 400~K in Fig.~A(b)] as well as for the main gradient-spacer sample series (main text) merge, which shows that the remaining offset is of non-exchange origin. We attribute this offset to magneto-static in nature orange-peel coupling, independent of the details of the spacer, and subtract it from the $H_\mathrm{c}$ data in the Discussion of the main text.

%##################### Figure B #####################
%===================================================
\begin{figure*}%[!b]
\includegraphics[width=16cm]{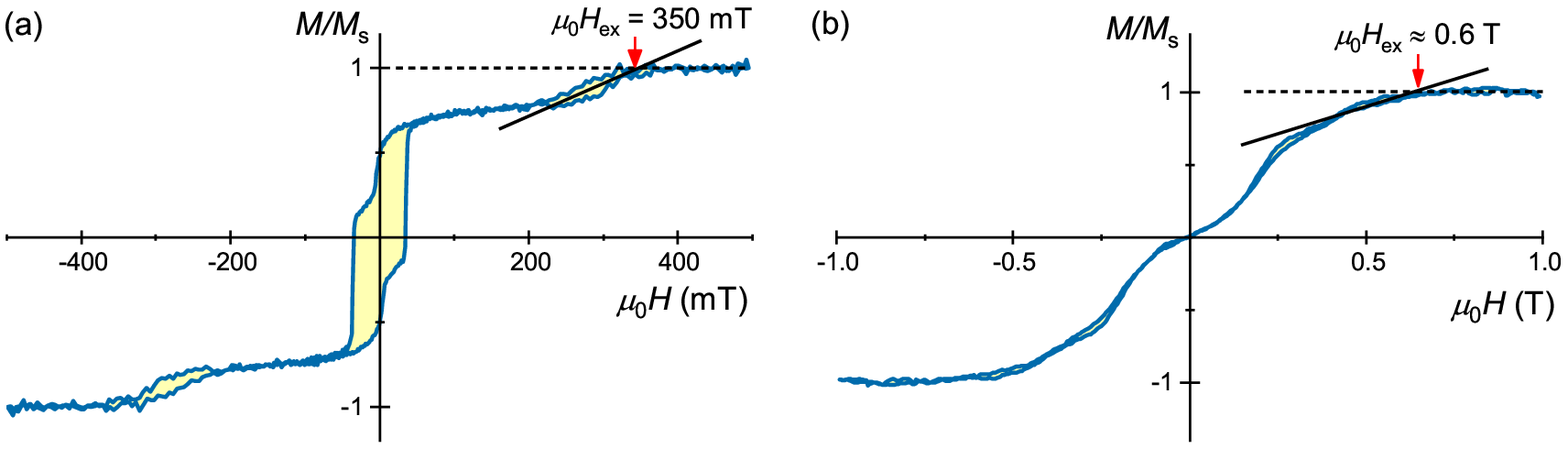}\\
\textbf{Figure B.} (a) Normalized magnetization vs field for sample of uniform-spacer series with spacer Cr(1)/Fe$_{25}$Cr$_{75}$(1)/Cr(1) and Fe(8)/Cr(1.2)/Fe(2) synthetic ferrimagnet as biasing layer, F$_\mathrm{pin}$. (b) Normalized magnetization vs field for symmetric trilayer sample with pure-Cr spacer in strong RKKY regime, Fe(2)/Cr(1.2)/Fe(2). RKKY exchange fields are defined as saturation fields taken as crossing points of linearly interpolated $m(H)$ prior and after saturation (solid and dashed lines). The data in both (a) and (b) were taken at room temperature.
\label{fig_B}
\end{figure*}
%===================================================

\subsection*{Note 2. Determining strength of RKKY interaction}

In order to estimate the effective RKKY exchange fields acting on the inner dilute ferromagnetic spacer Cr(1)/Fe$_{25}$Cr$_{75}$(1)/Cr(1), we first determine the strength of the antiferromagnetic RKKY interaction (RKKY exchange constant $J$) at the distance of $d_\mathrm{Cr} \approx$~1.2~nm from the Fe interfaces. This can be done by comparing the following two multilayer systems, which is very illustrative as well as conclusive in terms of the general calibration of the experiment. The first system is the SFM, Fe(8)/Cr(1.2)/Fe(2), used as the pinned layer in the main sample series. The second system is a separately fabricated symmetric Fe(2)/Cr(1.2)/Fe(2) trilayer. The magnetization curves for the two structures, shown respectively in panels (a) and (b) of Fig.~B, exhibit high saturation fields, which indicate a strong antiferromagnetic RKKY coupling. With no in-plane magnetic anisotropy in the films, verified by separate FMR vs in-plane-angle measurements (not shown), the standard method of defining the saturation field in the measured $M$-$H$ straightforwardly yields the exchange fields of the RKKY coupling of 0.35 and 0.6~T, for the two samples in Figs.~B(a) and (b), respectively. Using these $H_\mathrm{ex}$, the saturation magnetization of Fe, $M_\mathrm{Fe} \approx 1.7\cdot 10^6$~A/m, and the relevant thicknesses of the Fe layers, equation (2) of the main text yields $J_\mathrm{RKKY} \approx -1.0$~mJ/m$^2$, which, as expected, is equal for both cases (a, b).

\end{document}